%Paper: cond-mat/9407073
%From: SODANO@pg.infn.it
%Date: Sat, 16 JUL 94 17:08 GMT

\input harvmac
\Title{\vbox{\baselineskip12pt\hbox{MPI-PhT/94-42}\hbox{DFUPG 91/94}}}
{\vbox{\centerline {Oblique Confinement and Phase Transitions}
\vskip2pt\centerline{in Chern-Simons Gauge Theories}}}
\centerline{M. C. Diamantini and P. Sodano}
\medskip
\centerline{I.N.F.N. and Dipartimento di Fisica, Universit\`a di Perugia}
\centerline{via A. Pascoli, I-06100 Perugia, Italy}
\bigskip
\centerline{C. A. Trugenberger}
\medskip
\centerline{Max-Planck-Institut f\"ur Physik}
\centerline{F\"ohringer Ring 6, D-80805 M\"unchen, Germany}
\vskip .3in
\noindent
We investigate non-perturbative features of a planar Chern-Simons
gauge theory modeling the long-distance physics of quantum Hall
systems, including a finite gap $M$ for excitations. By
formulating the model on a lattice we identify the relevant
topological configurations and their interactions. For $M >
M_{cr}$, the model exhibits an {\it oblique confinement} phase,
which we identify with Laughlin's incompressible quantum fluid. For
$M < M_{cr}$, we obtain a phase transition to a {\it
Coulomb } phase or a {\it confinement} phase, depending on the
value of the electromagnetic coupling.

\Date{July 1994}
Planar gauge fields play an important role in effective field theories
describing the low-energy degrees of freedom of two-dimensional
condensed matter systems \ref\fra{For
a review see: E. Fradkin, "Field Theories of Condensed Matter Systems",
Addison-Wesley, Reading (1991).} .
When the discrete $P$ and $T$ symmetries are either explicitly or
spontaneously broken, the dynamics of the gauge fields is usually
governed by the topological Chern-Simons term. In particular, the
theory with Lagrangian (we use units $c=1$, $\hbar =1$)
\eqn\lag{{\cal L} = {\kappa \over \pi}\ A_{\mu }\epsilon ^{\mu \alpha
\nu }\partial _{\alpha }B_{\nu } + {\eta \over \pi } \ B_{\mu }
\epsilon ^{\mu \alpha \nu } \partial _{\alpha }B_{\nu } }
has been proposed \ref\fwz{J. Fr\"ohlich
and A. Zee, Nucl. Phys. B364 (1991) 517;
X.-G. Wen and A. Zee, Phys. Rev. B46 (1993) 2290.} \ as the effective field
theory describing the long distance physics of {\it chiral incompressible
fluids}. Here, the current
$j^{\mu } \equiv {\kappa \over \pi }\ \epsilon ^{\mu \alpha \nu
} \partial _{\alpha }B_{\nu } $
describes matter fluctuations of charge $\kappa $
above the ground state. The first term
in \lag \ is the standard electromagnetic coupling, while the second
term describes the kinetic term for matter fluctuations.
This can be written as a non-local Hopf interaction; the introduction
of the effective pseudovector
gauge field $B_{\mu }$ allows however to avoid
non-local terms in the effective field theory \ref\zee{For a review see:
A. Zee, "Long Distance Physics of Topological Fluids", Progr. Theor.
Phys. Supp. 107 (1992) 77.} .
For $\eta =$ even integer,
the effective field theory \lag \ describes the long-distance physics
of {\it chiral spin liquids}
\ref\wwz{X.-G. Wen, F. Wilczek and A. Zee, Phys. Rev. B39 (1989) 11413.} ;
in this case, the $P$ and $T$ symmetries are
spontaneously broken. For $\eta =$ odd integer, the same theory
describes the long-distance physics of {\it Laughlin's incompressible quantum
fluids},
which are the matter ground states at the plateaus of the quantum
Hall effect \ref\gip{For a review see: "The Quantum Hall Effect", R. E. Prange
and S. M. Girvin eds., Springer-Verlag, New York (1990).} .
In this case the $P$ and $T$ symmetries are explicitly broken
by the external magnetic field and $1/\eta $ plays the role of the
{\it filling fraction} \ref\sos{See also: G. W.
Semenoff and P. Sodano, Phys. Rev. Lett. 57 (1986) 1195.} \ as
can be easily recognized by integrating out the matter degrees of
freedom and computing the current induced by a constant, uniform
electric field.

Up to now, only perturbative analyses of the model \lag \ are available.
Vortices have been coupled to the system only as external sources, in order
to classify their quantum numbers \fwz . In this paper we do two things.
First we enlarge the model \lag \ by adding to it the three possible
terms of dimension $[{\rm mass}^4]$ coupling the dual field strengths
$F^{\mu }\equiv \epsilon ^{\mu \alpha \nu }\partial _{\alpha } A_{\nu }$
and $f_{\mu }\equiv \epsilon ^{\mu \alpha \nu }\partial _{\alpha } B_{\nu }$:
\eqn\oum{{\cal L}=-{1\over 2e^2}F_{\mu }F^{\mu } + {\kappa \over \pi}
A_{\mu }\epsilon ^{\mu \alpha \nu }\partial _{\alpha } B_{\nu }-\lambda
F_{\mu }f^{\mu }-{1\over 2g^2}f_{\mu }f^{\mu }+{\eta \over \pi } B_{\mu }
\epsilon ^{\mu \alpha \nu }\partial _{\alpha }B_{\nu } \ .}
These can be interpreted as the next-to-leading terms appearing in a
derivative expansion of a relativistic, gauge invariant effective action
for the incompressible quantum fluids. They provide {\it dynamics} for the
gauge
fields $A_{\mu }$ and $B_{\mu }$, which become propagating degrees of
freedom. We stress that this dynamics is not meant to reproduce exactly
the dynamics of matter fluctuations about incompressible quantum fluids;
rather it is meant to incorporate one essential feature of this dynamics
which is not described by \lag ,
namely the existence of a {\it finite gap} $M$ for the
excitations.

Secondly, we probe {\it non-perturbative features} of our model, by formulating
the corresponding Euclidean action on a lattice.
Indeed, the two Abelian gauge symmetries of \oum \ have to be considered
as {\it compact} $U(1)$ symmetries. The compactness of the gauge groups
leads to the existence of {\it topological excitations} \ref\pol{For a
review see: A. Polyakov, "Gauge Fields and Strings", Harwood Academic,
London (1987).},
which can drive {\it phase transitions}
\ref\sav{For a review see A. Savit, Rev. Mod. Phys. 52 (1980) 453; H.
Kleinert, "Gauge Fields in Condensed Matter Physics", World
Scientific, Singapore (1989).} .

Let us first describe the additional terms in \oum . The $F_{\mu }F^{\mu }$
term is the standard Maxwell term of (2+1)-dimensional electrodynamics.
The corresponding coupling constant $e^2$ has dimension [mass]: in modeling
quantum Hall systems it can be viewed as the fundamental length scale set
by the external magnetic field, namely $\alpha /\ell $, with $\alpha $ the
fine structure constant and $\ell $ the magnetic length.
The $f_{\mu }f^{\mu }$ term represents a
current-current interaction term for the
matter: it produces a mass $2\eta g^2/\pi $ for free matter fluctuations, the
so called magnetophonons \ref\mag{S. M. Girvin, "Collective Excitations",
in \gip .} .
Finally, the $F_{\mu }f^{\mu }$ term can be viewed (upon an integration
by parts) as an electromagnetic coupling of the matter vorticity
$\epsilon ^{\mu \alpha \nu }\partial _{\alpha }j_{\nu }$.
In our model, the coupling constant $\lambda $ is fixed as follows.  By
integrating out the electromagnetic gauge field $A_{\mu }$, we obtain
an effective theory for the matter degrees of freedom: \eqn\etm{{\cal
L}^B_{\rm eff}=-{1\over 2} \left( {1\over g^2}- e^2  \lambda ^2
\right) \ f_{\mu }f^{\mu } + {e^2\kappa ^2 \over 2 \pi ^2} B_{\mu }
\left( \delta ^{\mu \nu }- {\partial ^{\mu }\partial ^{\nu } \over
\Delta }   \right) B_{\nu } + {{\eta - \kappa \lambda e^2 } \over
\pi } B_{\mu } \epsilon ^{\mu \alpha \nu }\partial _{\alpha }B_{\nu }
\ .} For generic $\lambda $,
this theory contains both a Higgs mass and a topological Chern-Simons mass.
We shall fix $\lambda $ by the requirement that the induced Chern-Simons
term cancels exactly the bare one, so that only the Higgs mass survives:
\eqn\del{\lambda = {\eta \over \kappa e^2} \ .}
As we shall see below, it is exactly this requirement which allows
a consistent formulation of the model \oum \ on the lattice.

When \del \ is satisfied, \etm \ describes a parity doublet of excitations
with spin $\pm 1$ \ref\bin{B. Binegar, J. Math. Phys. 23 (1982) 1511.} \
and mass
\eqn\mas{M = {{e g \kappa \over \pi} \over \sqrt{1-{\eta ^2 g^2
\over \kappa ^2 e^2}}} \ .}
This mass represents the effective gap for matter
excitations, when electromagnetic fluctuations
are taken into account. The mass scale $g^2$ can be
expressed in terms of the fundamental mass scale $e^2$ by introducing
a dimensionless parameter. By choosing
$g^2=[\pi ^2x^2/(1+\pi ^2x^2)] \kappa ^2 e^2/\eta ^2$ the gap becomes
$M = x \kappa ^2 e^2/\eta $
and has the correct scaling with charge and
filling fraction \ref\tsu{R. R. Du, H. L. Stormer, D. C. Tsui, L. N. Pfeiffer
and K. W. West, Phys. Rev. Lett. 70 (1993) 2944.} \ to model the
quantum Hall gap.
Here $x$ can be thought of as a phenomenological parameter encoding all
microscopic effects which have an influence on the gap, like the range
and strength
of the inter-particle interaction and the amount of impurities.
With this choice of $g^2$,
the original model \lag \ is recovered in the limit $e^2 \to \infty $.

In the following we shall study the {\it phase structure} of our
model.
To this end we study the {\it topological excitations} due to the
compactness of the two Abelian gauge symmetries of \oum .
We introduce a cubic lattice with lattice spacing $l$ and lattice sites
denoted by ${\bf x}$, on which we define
the following forward and backward lattice derivatives and shift operators:
\eqn\lds{\eqalign{d_{\mu } f({\bf x}) &\equiv {{f({\bf x} +\hat \mu
l) - f({\bf x})}\over l}\ , \qquad S_{\mu } f({\bf x}) \equiv f({\bf
x}+\hat \mu l) \ ,\cr
\hat d_{\mu }f({\bf x}) &\equiv
{{f({\bf x})-f({\bf x}-\hat \mu l) }\over l} \ , \qquad
\hat S_{\mu } f({\bf x}) \equiv f({\bf x}-\hat \mu l) \ ,\cr }}
where $\hat \mu $ denotes a unit vector in direction $\mu $. To each
link $\{ {\bf x}, \mu \}$ of the lattice we assign two real gauge fields
denoted by $A_{\mu }({\bf x})$ and $B_{\mu }({\bf x})$. These are compact
variables defined on a circle of radius $2\pi /l $:
$-(\pi /l)< A_{\mu }, B_{\mu }<(\pi /l) $. This definition fixes
the normalization of the gauge fields; consequently, no coupling
constant in \oum \ can be reabsorbed in a redefinition of $A_{\mu }$
and $B_{\mu }$.

In order to formulate an Euclidean version of \oum \ on the lattice, we have
to face the problem of defining a lattice version of the (Euclidean)
Chern-Simons
operator $\epsilon _{\mu \alpha \nu } \partial _{\alpha } $, which can be
viewed as the square root of the familiar Maxwell operator.  It turns out that
on a cubic lattice there is no gauge invariant local operator whose square
reproduces the Maxwell operator. Rather , one can define the following two
lattice operators \ref\frm{The operator $K_{\mu \nu}$ was first defined in
J. Fr\"ohlich and P. Marchetti, Comm. Math. Phys. 121 (1989) 177; see also
D. Eliezer and G. W. Semenoff, Ann. Phys. (N.Y.) 217 (1992) 66.} :
\eqn\lcs{K_{\mu \nu} \equiv S_{\mu } \epsilon _{\mu \alpha \nu} d_{\alpha }\ ,
\qquad \hat K_{\mu \nu} \equiv \epsilon _{\mu \alpha \nu} \hat d_{\alpha } \hat
S_{\nu } \ ,}
where there is no summation over equal indices. These operators are gauge
invariant,
\eqn\gik{K_{\mu \nu }d_{\nu } = \hat d_{\mu } K_{\mu \nu }
=0\ ,\qquad
\hat K_{\mu \nu } d_{\nu } = \hat d_{\mu } \hat K_{\mu \nu }
=0 \ ,} and their product reproduces the Maxwell operator,
\eqn\prm{K_{\mu \alpha }\hat K_{\alpha \nu } = \hat K_{\mu \alpha }
K_{\alpha \nu } = -\delta _{\mu \nu } \nabla ^2  + d_{\mu }\hat d_{\nu }\ ,}
where $\nabla ^2 \equiv \hat  d_{\mu } d_{\mu }$ is the three-dimensional,
Euclidean Laplace operator on the lattice.  Note also that the two operators
$K_{\mu \nu }$ and $\hat K_{\mu \nu }$ are interchanged upon a
summation by parts on the lattice.

In order to take into account the periodicity of the link variables $A_{\mu }$
and $B_{\mu }$ we introduce four sets of integer link variables and we posit
the following Euclidean model of the Villain type \sav :
\eqn\lam{\eqalign{Z &= \sum _{{\{n_{\mu }\}, \{l_{\mu }\} }\atop {
\{m_{\mu }\} , \{ k_{\mu }\} }} \int _{-\pi \over l}^{\pi \over l}
 {\cal D} A_{\mu } {\cal D} B_{\mu } \ {\rm exp}(-S) \ ,\cr
S &= \sum_{{\bf x}, \mu }\  {l^3\over 2e^2} \left( F_{\mu } +{2\pi
\over l^2} n_{\mu } \right) ^2 - i{l^3 \kappa \over 2\pi } \left(
A_{\mu } + {2\pi \over l} l_{\mu } \right) K_{\mu \nu } \left( B_{\nu
} +{2\pi \over l} m_{\mu } \right) \cr  &-i{l^3 \kappa \over 2\pi }
\left( B_{\mu }+ {2\pi \over l} m_{\mu } \right) \hat K_{\mu \nu }
\left( A_{\nu }+{2\pi \over l} l_{\nu } \right) \cr
&+{l^3 \eta \over \kappa e^2} \left( F_{\mu } +{2\pi \over l^2}
n_{\mu } \right) \left( f_{\mu } +{2\pi \over l^2} k_{\mu } \right)
+{l^3\over 2g^2} \left( f_{\mu } + {2\pi \over l^2} k_{\mu } \right) ^2
\cr &-i{l^3\eta \over 2\pi } \left( B_{\mu } +{2\pi \over l} m_{\mu
}\right) \left( K_{\mu \nu }+\hat K_{\mu \nu } \right) \left( B_{\nu }
+ {2\pi \over l} m_{\nu } \right) \ ,\cr }}
where we have introduced
the notation ${\cal D} A_{\mu } \equiv \prod _{{\bf x}, \mu } dA_{\mu
}({\bf x}) $ and we define the lattice dual field strengths as
$F_{\mu } \equiv K_{\mu \nu }A_{\nu }$ and
$f_{\mu }\equiv K_{\mu \nu }B_{\nu }$.
Due to the property \prm , the terms $\sum_{{\bf x},
\mu } F_{\mu }^2$ and $\sum_{{\bf x}, \mu } f_{\mu }^2$ reproduce
the familiar lattice Maxwell action.
The partition function \lam \ is clearly invariant under
shifts $A_{\mu }\to A_{\mu }+ 2\pi i_{\mu }/l$ and $B_{\mu } \to
B_{\mu } +2\pi j_{\mu }/l$ with integer $i_{\mu }$ and $j_{\mu }$,
since these can be reabsorbed by a redefinition of the integer
link variables $n_{\mu }$, $m_{\mu }$, $l_{\mu }$ and $k_{\mu }$.

Gauge invariance, instead requires the {\it quantization} of the
parameters $\kappa $ and $\eta $. The
boundary conditions are such that the dual field strengths $F_{\mu }$
and $f_{\mu }$ vanish modulo $2\pi /l^2$ at  infinity. Consider now a
gauge transformation $A_{\mu }\to A_{\mu } + d_{\mu }\Lambda $ which
wraps non-trivially around one of the three directions, say $\Lambda
(-\infty, x^2, x^3)=0$, $\Lambda (+\infty, x^2, x^3) =2\pi n$
with $n$ an integer. Under such a gauge transformation the lattice
action \lam \ changes by  the surface term obtained by summing by
parts the second term in \lam :
\eqn\cgt{\Delta S = \sum_{x^2, x^3}
-i\kappa n \  t(+\infty, x^2, x^3) \ ,}
where $t\equiv 2\pi lK_{1\nu} m_{\nu}$ is an integer multiple
of $2\pi $. Gauge invariance requires that $\Delta S$ vanishes
modulo $i2\pi $. This is realized when the coupling constant
$\kappa $ satisfies an integer quantization condition.
Correspondingly, gauge invariance under topologically non-trivial
gauge transformations $B_{\mu }\to B_{\mu }+ d_{\mu }\Lambda $
requires the integer quantization of the coupling constant $\eta $:
$k=p \in Z$, $\eta =q \in Z$. Thus, the quantization of the inverse
filling fraction $1/\eta $ is a consequence of the compactness
of the effective gauge field $B_{\mu }$.

We now rewrite \lam \ in a fashion which exposes explicitly
the topological configurations and their interactions. To this
end we decompose $n_{\mu }$ and $k_{\mu }$ as
$n_{\mu } \equiv lK_{\mu \nu } l_{\nu } +
a_{\nu }$,
$k_{\mu } \equiv lK_{\mu \nu } m_{\nu } + b_{\nu }$,
with $a_{\mu }$ and $b_{\mu }$ integers. Correspondingly, the
sum over all configurations $\{ n_{\mu }\} $ and $\{ k_{\mu }\} $
in \lam \
can be replaced by a sum over all configurations $\{a_{\mu }\}$ and
$\{b_{\mu }\}$. By changing variables $A_{\mu }\to A_{\mu } +
(2\pi /l) l_{\mu }$ and $B_{\mu }\to B_{\mu }+(2\pi /l) m_{\mu }$
in the integration and performing the sum over all
configurations $\{l_{\mu }\}$ and $\{m_{\mu }\}$ we obtain
\eqn\its{\eqalign{Z &= \sum_{{\{a_{\mu }\}} \atop {\{b{\mu }\}}}
\int _{-\infty }^{+\infty } {\cal D} A_{\mu } {\cal D} B_{\mu } \
{\rm exp}(-S)\ ,\cr
S &= \sum_{{\bf x}, \mu }\  {l^3\over 2e^2} F_{\mu }^2 -i{l^3 p
\over \pi } A_{\mu }K_{\mu \nu }B_{\nu }+{l^3q \over e^2p}
F_{\mu }f_{\mu }+{l^3\over 2g^2} f_{\mu }^2 -i{l^3q \over \pi}
B_{\mu }K_{\mu \nu }B_{\nu } \cr
&+{2\pi ^2\over le^2}\  a_{\mu }^2 +{2\pi ^2\over lg^2}\  b_{\mu }^2
+{4\pi ^2 q \over lp e^2}\  a_{\mu }b_{\mu } \cr
&+ {2\pi l\over e^2} A_{\mu } \hat K_{\mu \nu } \left( a_{\nu }
+{q\over p} b_{\nu }\right)
+{2\pi l\over e^2} B_{\mu } \hat K_{\mu \nu }
\left( {e^2\over g^2} b_{\nu }
+{q \over p
} a_{\nu }\right) \ .\cr }}
Finally, we perform the Gaussian integrations over
$A_{\mu }$ and $B_{\mu }$. Here the condition \del \ is crucial:
indeed, for other values of $\lambda $ the quadratic kernels
in the Gaussian integrals would not be invertible. The final
result takes the form $Z=Z_0\ Z_{Top}$, where $Z_0$ is the lattice
partition function for the non-compact, Euclidean version of \oum \
and $Z_{Top}$ is given by
\eqn\top{\eqalign{Z_{Top}=\sum_{{\{a_{\mu }\}} \atop {\{b_{\mu }\}}}
{\rm exp} \sum_{{\bf x}, \mu } &-{2\pi ^2\over le^2} \left(
a_{\mu }+{q\over p}b_{\mu }\right) {{M^2 \delta _{\mu \nu }
-d_{\mu }\hat d_{\nu }} \over {M^2-\nabla ^2}} \left(
a_{\nu }+{q\over p}b_{\nu } \right) \cr
&-{2e^2 p^2 \over lM^2}\  b_{\mu }{{M^2
\delta _{\mu \nu }-d_{\mu }\hat d_{\nu }}\over {M^2-\nabla ^2}}
b_{\nu }-i{4\pi p \over l}\  b_{\mu }
{{\hat K_{\mu \nu }}\over {M^2-\nabla ^2}} \left(
a_{\nu }+{q\over p}b_{\nu } \right)\ .\cr}}

The integer link variables $a_{\mu }$ and $b_{\mu }$ form
the topological configurations of the model. They arise as
the integer parts of the dual field strengths $F_{\mu }$ and
$f_{\mu }$, respectively. Therefore the corresponding topological
excitations can be interpreted as {\it magnetic flux strings}
and {\it quasi-particle current strings}. The quasi-particles
represent localized, collective matter excitations and have
to be distinguished from the original matter particles of the
underlying microscopic model. The strings can be closed (rings), in
which case $\hat d_{\mu }a_{\mu }=0$ and $\hat d_{\mu }b_{\mu }=0$,
or open, in which case the integers
$l\hat d_{\mu } a_{\mu }$ and $l\hat d_{\mu }b_{\mu }$ represent
the {\it monopoles} corresponding to the two compact
Abelian gauge symmetries of \lam . In our Euclidean formalism,
these monopoles describe tunneling events corresponding
to the formation (or destruction) of magnetic fluxes and
quasi-particles.

Eq. \top \ describes the interactions between these topological
excitations. Note that magnetic flux strings appear only in the
combination $[a_{\mu }+(q/p)b_{\mu }]$. Due to the mass gap
$M$, all interactions are short-range; it is therefore a
good approximation to neglect the off-diagonal terms in the
interaction kernels and to assign a free energy
\eqn\fre{F = \left\{ {2\pi ^2\over le^2} \left( a+{q\over p}b
\right) ^2 +{2 e^2 p^2\over lM^2}\  b^2 -\mu \right\} N }
to a string of length $L=lN$ carrying magnetic and
quasi-particle quantum numbers $a$ and $b$. The last term in \fre \
represents the entropy of the string: the parameter $\mu $ is
given roughly by $\mu ={\rm ln} 5$, since at each step the string
can choose between 5 different directions. In a dilute instanton
approximation, in which all values $a_{\mu }, b_{\mu }\ge 2$ are
neglected it can be prooved that the correct value of $\mu $ is
the same for open and closed strings \ref\eis{M. B.
Einhorn and R. Savit, Phys. Rev. D19 (1979) 1198.} . In \fre \ we have
neglected all subdominant functions of $N$, like a ${\rm ln}N$
correction to the entropy and a constant term due to the monopole
contribution to the energy for open strings. Moreover, we have
neglected the imaginary term in the action \top . This is justified
self-consistently, since the contribution of this term vanishes
in all phases of the model, as we now show.

Long strings with quantum numbers $a$ and $b$ condense in the
ground state if the coefficient of $N$ in \fre \ is negative.
When two or more condensates are possible, one has to choose the
one with the lowest free energy. The condensation condition
describes the interior
of an ellipse with semi-axes $le^2 \mu /2\pi ^2$ and
$lM^2 \mu /2e^2 p^2$ on a non-rectilinear lattice of
magnetic and quasi-particle charges.
For generic values of $p$ and $q$
the phase structure may be quite complex, displaying various
{\it oblique confinement}
phases \ref\car{G.'t Hooft, Nucl Phys.
B190 [FS3] (1981) 455;
J. L. Cardy and E. Rabinovici, Nucl. Phys.
B205 [FS5] (1982) 1; J. L. Cardy, Nucl. Phys. B205 [FS5] (1982) 17.} .
Here we discuss the simpler case of
matter of charge $p$=1. In this case we find the following
phase structure:
\eqn\phs{\eqalign{le^2\mu < 2\pi ^2 &\to \cases{{M^2\over 2e^4}=
{x^2\over 2q^2}>
{1\over le^2\mu }\ , &oblique confinement\ ;\cr
{M^2\over 2e^4}=
{x^2\over 2q^2}<{1\over
le^2\mu }\ ,&Coulomb\ ;\cr} \cr le^2\mu > 2\pi ^2 &\to
\cases{{M^2\over 2e^4}=
{x^2\over 2q^2} > {1\over 2\pi ^2}\ , &oblique confinement\ ;\cr
{M^2\over 2e^4}=
{x^2\over 2q^2}<{1\over 2\pi ^2}\ ,&confinement\ .\cr} \cr }}
In the oblique confinement phase the ground state consists of
a condensate of strings carrying magnetic flux $a=\pm q$ and
quasi-particle number $b=\mp 1$.
The confinement phase is
characterized by a ground state consisting of a condensate of
magnetic flux strings. In the Coulomb phase, instead,
there is no condensation
of topological excitations in the ground state.

The order parameter distinguishing the various phases is the
{\it Wilson loop} \pol \  expectation value in the
corresponding ground states. We shall present the details
of this computation elsewhere \ref\dst{M. C. Diamantini, P.
Sodano and  C. A. Trugenberger, in preparation.} .
Here we report only the results.
In the oblique
confinement phase,
the only free (non-confined) excitations
carry magnetic and quasi-particle quantum numbers in the ratio
$a/b=-q$. A magnetic flux must consequently carry {\it fractional
quasi-particle number} $-1/q$. This composite object is therefore an
{\it anyon} \fra \ excitation with {\it fractional statistics}.
The free excitations in the confinement phase
are magnetic fluxes: in this phase quasi-particles are confined.
In the Coulomb phase, both magnetic
fluxes and quasi-particle constitute free excitations.

It is known \ref\gmr{S. M. Girvin and A. H. MacDonald, Phys.
Rev. Lett. 58 (1987) 1252; N. Read, Phys. Rev. Lett. 62 (1989) 86.}
\ that the microscopic Laughlin wave functions
\ref\lau{R. B. Laughlin, "Elementary
Theory: The Incompressible Quantum Fluid", in \gip .}\ for
the incompressible quantum fluids describe a state
in which an odd number of statistical fluxes are bound to
the electrons. This fact is at the basis of Jain's theory
\ref\jai{For a review see: J. K. Jain, Adv. in Phys. 41 (1992) 1.}\
of composite electrons and of most field theoretic treatments
\ref\flo{S. C. Zhang, T. Hansson and S. Kivelson, Phys. Rev.
Lett. 62 (1989) 82; A. Lopez and E. Fradkin, Phys. Rev. B44
(1991) 5246, Phys. Rev. Lett. 69 (1992) 2126, Nucl. Phys. B
(Proc. Supp.) 33C (1993) 67.}\ of the quantum Hall effect.
We are thus led to identify the oblique confinement phase of
our model (for odd q) with the incompressible quantum fluid phase
of quantum Hall systems. Note however, that our mechanism of
condensation of composite objects is different from the usual one,
in which fictitious, statistical flux is attached to the physical
electrons. In our model the dual mechanism takes place: physical
magnetic flux is attached to collective
quasi-particle excitations and the so formed composite objects
condense in the ground state. Moreover, in previous field
theoretic formulations of the quantum Hall effect, flux is
attached to the electrons by the introduction of a Chern-Simons
term for the statistical gauge field. This is a
kinematical mechanism, since it follows from the Chern-Simons
Gauss law constraint. In our model, instead, the condensation of
composite objects is a {\it dynamical} mechanism. Correspondingly
we obtain the flux-unbinding phase transition to a Coulomb or to
a confinement phase, depending on the value of the electromagnetic
coupling
$le^2\mu $. This transition takes place when the parameter
$x/q$ diminishes under a critical value (depending on $le^2\mu $)
i.e. when either the phenomenological
gap parameter $x$ or the filling fraction $1/q$ become too small.

\listrefs
\end